\documentclass[%
 reprint,
 amsmath,amssymb,
 aps,
]{revtex4-1}

\usepackage{graphicx}
\usepackage{dcolumn}
\usepackage{bm}

\begin{document}

\preprint{APS/123-QED}

\title{Waveform--dependent absorbing metasurfaces}

\author{Hiroki Wakatsuchi}
 \email{hirokiwaka@gmail.com}
\author{Sanghoon Kim}%
\author{Jeremiah J. Rushton}%
\author{Daniel F. Sievenpiper}%
 \email{dsievenpiper@eng.ucsd.edu}
\affiliation{%
 Applied Electromagnetics Group, Electrical and Computer Engineering Department, 
 University of California, San Diego, California 92093, USA
}%

\date{\today}

\begin{abstract}
We present the first use of a waveform--dependent absorbing metasurface for high--power pulsed surface currents. The new type of nonlinear metasurface, composed of circuit elements including diodes, is capable of storing high power pulse energy to dissipate it between pulses, while allowing propagation of small signals. Interestingly, the absorbing performance varies for high power pulses but not for high power continuous waves (CWs), since the capacitors used are fully charged up. Thus, the waveform dependence enables us to distinguish various signal types (i.e.\ CW or pulse) even at the same frequency, which potentially creates new kinds of microwave technologies and applications.
\end{abstract}

\maketitle


\hyphenpenalty=1000

Conventional absorbers \cite{knott2006radar, MunkBook} are independent of incoming power levels and therefore absorb not only destructive high power signals but also small signals necessary for antenna communications. These two effects can be decoupled by introducing nonlinearity into a periodic structure or metasurface \cite{kivshar2003nonlinear} such that the artificially engineered surface possesses a low power surface impedance that is different from its high power absorbing response. Furthermore, we can take advantage of the fact that high power microwave signals generally come from pulsed sources, and design a structure in which the high power microwave energy is rectified and stored as a static field in the surface, and then dissipated between pulses. This provides freedom in the choice of lossy components, allowing us to further decouple the high power absorption properties from the small signal behavior and also design the surface to respond specifically to short pulses.  

Many nonlinear metamaterials and metasurfaces derive their nonlinear properties from altering the resonant paths of induced electric currents with nonlinear media (e.g.\ through semiconductive substrates \cite{chen2008experimental,THzActiveMTMpadilla,driscoll2009memory,liu2012terahertz} or geometrical modifications induced by heating \cite{reconfigurableSRRs} or magnetic force \cite{lapine2011metamaterials,lapine2011magnetoelastic}). These differ from our nonlinear metasurfaces based on diodes or \emph{nonlinear circuits}. Diodes switch between nonconducting and conducting states depending on the applied voltage and therefore are capable of handling incoming signals differently depending on the magnitude. Because diodes rectify high power signals and convert much of the energy to a static field, our nonlinear metasurfaces perform differently from others that convert incoming signals into high order modes \cite{klein2006second,wang2009second} or exploit wave mixing techniques \cite{PhysRevB.78.113102}. 

\begin{figure}[t!]
\includegraphics[width=0.49\textwidth]{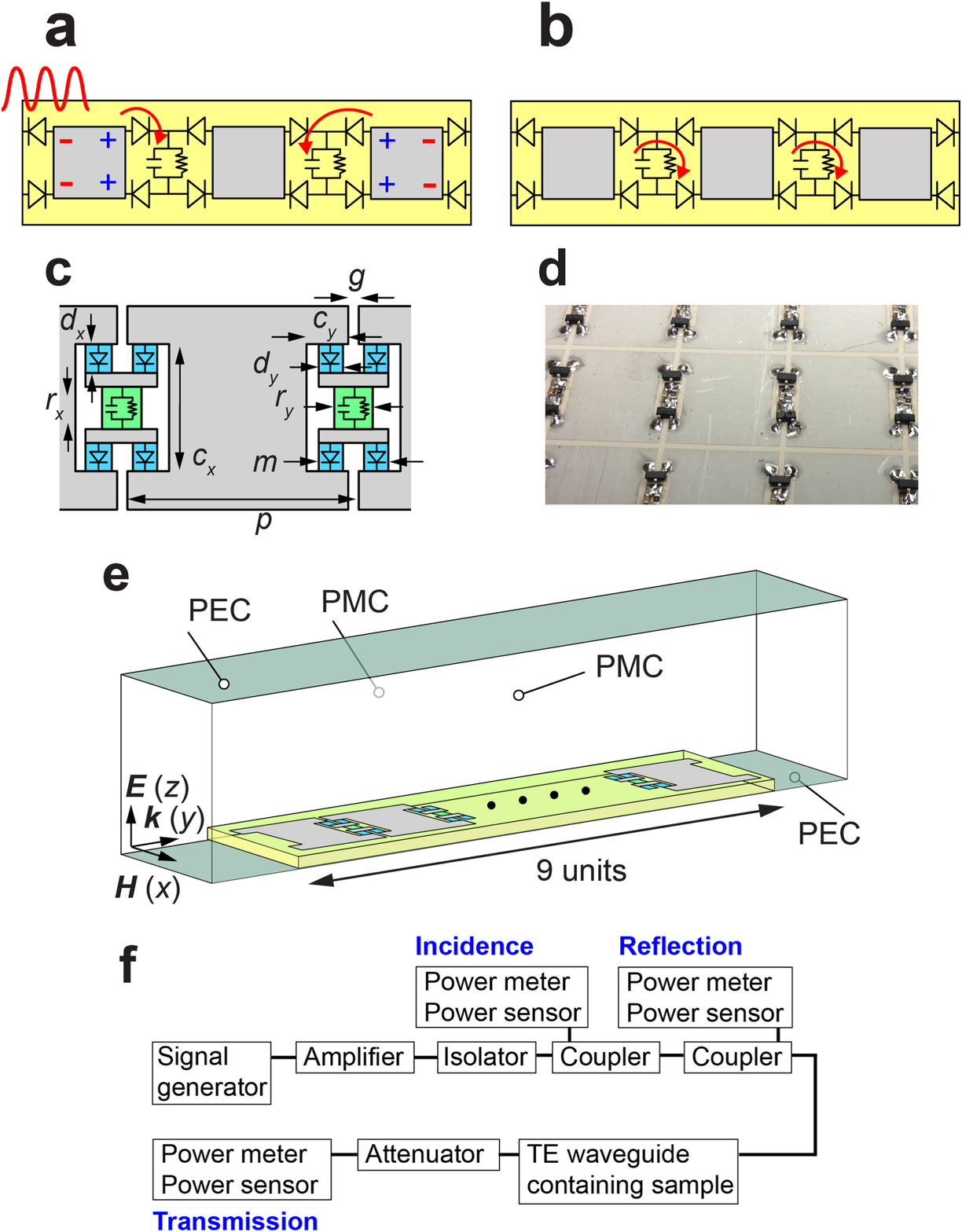}
\caption{\label{fig:models} Circuit--based nonlinear metasurface absorber. Response (a) during a pulse and (b) between pulses. The metasurface stores the energy of the high power pulse in capacitors during the illumination, while between pulses discharging the energy into resistors. (c) The geometry simulated. The periodic units were build on a 1.52 mm height dielectric substrate (Rogers 3003), which had a conductive ground plane on another side. The dimensions were given from $c_x=7.6$, $c_y=1.7$, $d_x=1.3$, $d_y=0.5$, $g=1.0$, $m=2.4$, $p=18.0$, $r_x=1.0$ and $r_y=2.0$ (all in mm). (d) Measurement sample. All the circuit components were soldered to the periodically metalized surface. (e) TEM waveguide simulation. The metasurface illustrated in (c) is placed at the bottom of the waveguide (22 mm tall and 18 mm wide). (f) Measurement system for pulse measurement. The number of periodic units deployed along the incident magnetic field was varied depending on the width of the TE waveguides used (WR284 or WR187) so that the structure fully occupied the bottom surface between the sidewalls. See Supplementary Material for more details on (e) and (f). }
\end{figure}

To understand how such a structure absorbs high power pulse energy consider Figs.\ \ref{fig:models} (a) and (b). Our metasurface is composed of circuit components (i.e.\ diodes, capacitors and resistors) as well as a conductive ground plane and periodic metal patches, separated by a dielectric substrate, Rogers 3003. During a pulse (i.e.\ Fig.\ \ref{fig:models} (a)) high power surface currents propagate on the patches. Once the voltage difference across a diode reaches the turn--on voltage, electric charges flow into a capacitor and the energy is stored. Between pulses (Fig.\ \ref{fig:models} (b)) the stored electric charges are dissipated in a resistor. Note that the metasurface does not respond to small signals and therefore behaves as an ordinary metal surface, since the voltage difference is not large enough to turn on the diodes. Although we have demonstrated absorbing behavior based on rectification previously \cite{aplNonlinearMetasurface,awplSievenpiper}, the new metasurface demonstrated here has three important differences. First, the previous metasurfaces contained vertical conducting vias, thus were not planar. In contrast, the structures designed here contain no vias, and are entirely printed on the top surface of the substrate. This may facilitate integration with other structures such as composite vehicle or aircraft skins. Second the new surface provides full--wave rectification using a four--diode bridge circuit, increasing the efficiency of microwave conversion into the static field (see Supplementary Material). Third, the new configuration has the parallel \textit{RC} circuit entirely isolated from the resonant structure by diodes. As a result, the self resonance frequency of the capacitors does not affect the low--power response of the surface. Without this feature, we would be unable to use sufficiently large capacitors to fully store the pulse energy without introducing unwanted behavior into the low--power absorption or scattering response. 

The metasurface in Figs.\ \ref{fig:models} (a) and (b) was realized using the printed structure shown in Fig.\ \ref{fig:models} (c). The structure was built on a 1.52 mm thick Rogers 3003 substrate. The periodicity of the printed metal pattern was 18 mm. A 10 k$\Omega$ resistor and 1 nF capacitor were attached at the gap between two small patches which were separated from the larger main patches by pairs of diodes forming a full wave rectification circuit. Note that the resistance value was set large enough to prevent significant direct dissipation from the diodes, although the suitable resistance value depends on several factors including the metasurface design and operating frequency. Capacitors were modeled to have a self resonant frequency at 0.3 GHz. Use of larger capacitance enables to store more energy. However, this delays the discharge process of the capacitors due to increase of the time constant. The operating frequency of the metasurface is scalable by changing the periodicity, as long as the switching speed of the diodes used is fast enough to respond to the incoming signal. More details on the structural configuration are summarized in Fig.\ \ref{fig:models} (c). Numerical simulations were performed by electromagnetic/circuit co--simulation in the same manner as our early work (see Ref.\ \cite{aplNonlinearMetasurface} and Supplementary Material). In this simulation we placed nine units of the periodic structure (i.e. 162 mm long in total) on the bottom of a TEM waveguide (22 mm tall and 18 mm wide).

The measurement sample is shown in Fig.\ \ref{fig:models} (d). The sample was fabricated to have the same conditions as the simulation model except for a few points. For example, the measurement needed to be performed with a finite width of TE waveguide (WR284 or WR187). Hence the measurement samples were designed to have a multiple number of periodic units for the direction of the incident magnetic field as well to fully occupy the bottom of the waveguides. Additionally, the measurement sample included schottky diodes (Avago; High Frequency Detector Diodes HSMS-2863/2864), which were included as a SPICE model in the simulation. The measurement was performed with two systems, respectively, for low power and high power measurements including high power pulses. For the low power measurement a network analyzer (Agilent E5071C) was connected to the two ends of a TE rectangular waveguide containing the sample. The measurement system used for the high power measurement is depicted in Fig.\ \ref{fig:models} (f). The details are described in Supplementary Material. 

\begin{figure*}[t!]
\includegraphics[width=0.95\textwidth]{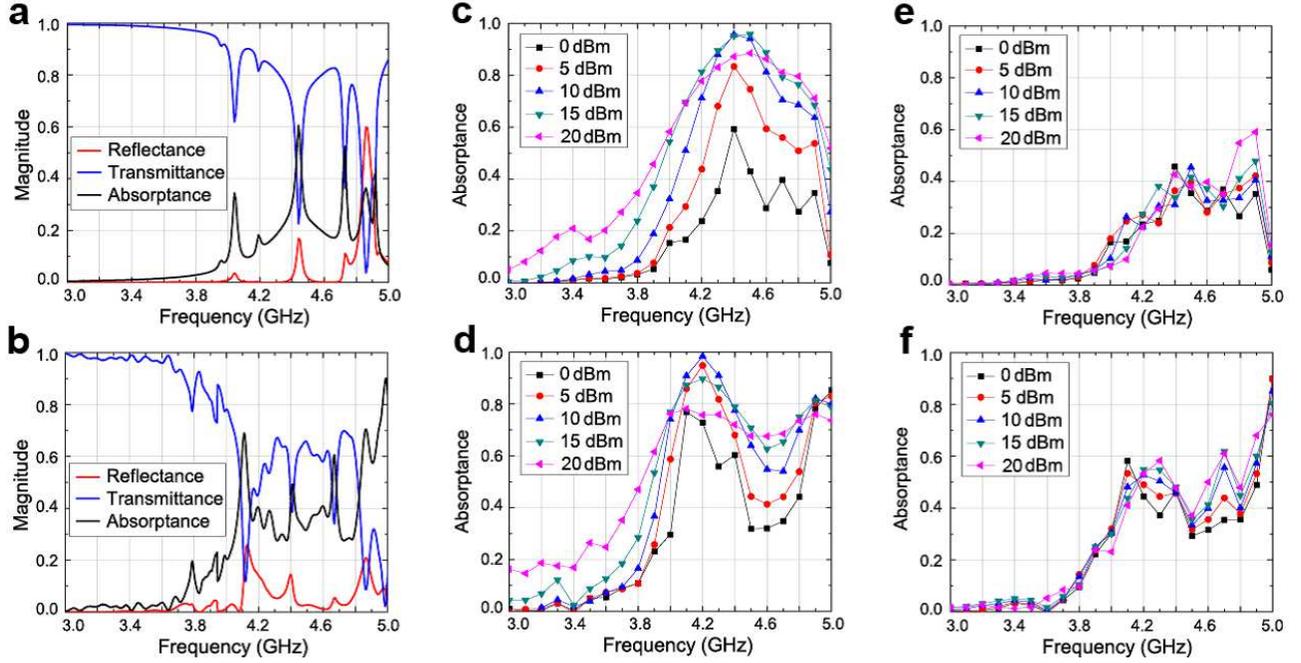}
\caption{\label{fig:highP} Scattering and absorbing profile. Simulation results are plotted in the left figures (i.e.\ (a), (c), and (e)), while measurement results in the right ((b), (d), and (f)). Low power CW responses are shown in (a) and (b). High power pulse absorptances are shown in (c) and (d), while high power CW absorptances in (e) and (f). In (c) and (d) the pulse width was set to 50 ns. }
\end{figure*}

Simulation results of the low power scattering and absorption profile are shown in Fig.\ \ref{fig:highP} (a). As seen in this figure, the metasurface mostly transmitted the low power signal with some limited absorption around 4.4 GHz. However, once the input signal was switched to 50 ns high power pulses, an absorptance peak emerged more strongly as plotted in Fig.\ \ref{fig:highP} (c). For instance, around 4.4 GHz the absorptance reached almost 100 \% with 10 to 15 dBm pulses. Further increasing the input power reduced the absorptance a little but resulted in relatively broadband behavior. Also note that at 4.2 GHz, where the low power transmittance was more than 80 \%, the absorptance was increased from below 20 \% to more than 80 \%. The behavior of the nonlinear metasurface changed from a relatively low loss surface to an absorptive surface due to the combination of rectification and the energy storage in the capacitors. A proof of the rectification to a static field is seen in Supplementary Material. Another point is that the absorption is reduced for CW signals as seen in Fig.\ \ref{fig:highP} (e). This is because rectification ceases once the capacitors are charged up to the level of the incoming signal. The high power CW performance is controllable by varying the resistance values (see Supplementary Material), i.e.\ more dissipation occurs by reducing the resistance value, which permits direct energy dissipation during rectification. In Fig.\ \ref{fig:highP} (e), however, we maximized the pulse effect, i.e.\ the contrast between high power pulse and high power CW, which is a distinctive feature of our nonlinear metasurface compared to other nonlinear metasurfaces (e.g.\ \cite{chen2008experimental,THzActiveMTMpadilla,driscoll2009memory}). 

The numerical results were then validated by measurement results. Fig.\ \ref{fig:highP} (b) shows almost the same characteristics as Fig.\ \ref{fig:highP} (a) but with a minor low frequency shift and slightly reduced transmittance around 4.4 GHz. These discrepancies can be mainly attributed to parasitic circuit parameters (especially in the diodes), which were not modeled in the simulation. However, the absorbing performance was still significantly enhanced as in the simulation, once the input source was switched to high power pulses (Fig.\ \ref{fig:highP} (d)). A strong absorptance peak was observed at 4.2 GHz with a 10 dBm pulse, while further increase of the input power resulted in small reduction but with broadband behavior. Also, at 4.0 GHz the metasurface absorbed 80 \% of 15 dBm pulse energy, while transmitting the same percentage of a low power signal. Thus, the measurements validate the simulations, and demonstrate that the surface is transformed from relatively low loss to highly absorbing under high power illumination. Also note that the absorbing performance was reduced when the input signal was changed from high power pulses to high power CW signals (Fig.\ \ref{fig:highP} (f)) as expected. The relatively large absorptance seen around 5.0 GHz is assumed to be due to parasitics of the diode packages. Therefore, the absorptance value remained the same with various input powers. 

\begin{figure}[t!]
\includegraphics[width=0.48\textwidth]{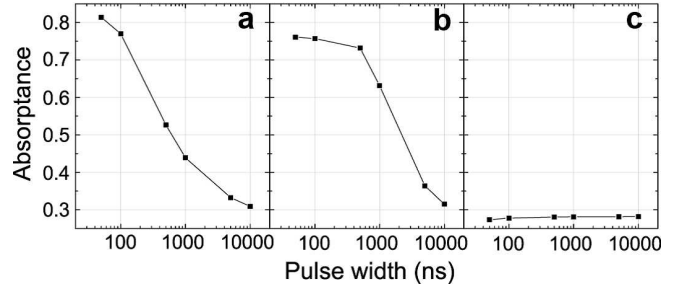}
\caption{\label{fig:pulseWidthDep} Absorptance dependence on pulse width. (a) Simulation result at 4.2 GHz and (b) measurement result at 4.0 GHz with a 15 dBm pulse. As the pulse width became longer, the absorbing performance decreased, approaching the high power CW performance in Figs.\ \ref{fig:highP} (e) and (f). (c) Simulation result at 4.2 GHz with a 15 dBm pulse without capacitors. The pulse width dependence disappeared by removing capacitors which play an important role to exhibit waveform dependence. }
\end{figure}

The transition between a 50 ns pulse and CW was more clearly investigated in Fig.\ \ref{fig:pulseWidthDep} (a). In this simulation the pulse width was varied from 50 ns to 10 $\mu$s. As seen in this figure the absorptance curve nearly saturated at 10 $\mu$s, which agrees with the time constant $\tau$ derived from the capacitance $C$ and resistance $R$ used, i.e.\ $\tau=RC=10$ k$\Omega \cdot 1$ ns$~=$ 10 $\mu$s. Also, the saturated absorptance at 10 $\mu$s (i.e.\ 0.31) was close to the high power CW performance (i.e.\ 0.28) (see Fig.\ \ref{fig:highP} (e)). Similar trends were observed in the measurement (see Fig.\ \ref{fig:pulseWidthDep} (b)). A minor delay of the saturation can be explained by some parasitic parameters in the circuit components, which increased the time constant slightly. However, the 10 $\mu$s absorptance (i.e.\ 0.32) was still in proximity with the CW result (i.e.\ 0.30) (see Fig.\ \ref{fig:highP} (f)). For these waveform dependences, capacitors play an important role as shown in Fig.\ \ref{fig:pulseWidthDep} (c), where the same simulation as Fig.\ \ref{fig:pulseWidthDep} (a) was performed but without capacitors (see Supplementary Material for more details such as low power and high power absorptance). Clearly, the pulse width dependence seen in Fig.\ \ref{fig:pulseWidthDep} (a) disappeared here, since there was no capacitor to temporarily store the rectified energy. In other words, capacitors are indispensable for creating the waveform dependence, which, as explained above, can be controlled by the resistance and capacitance used. 

\begin{figure}[t!]
\includegraphics[width=0.48\textwidth]{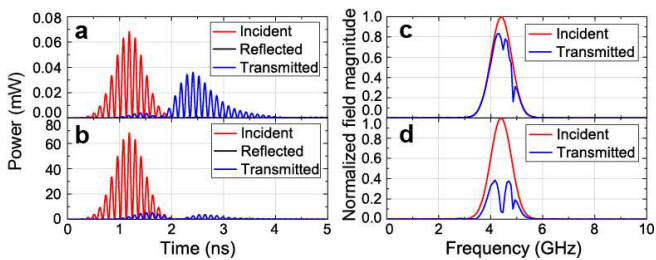}
\caption{\label{fig:gaussianPulse} Response to a Gaussian pulse modulated by a cosine function. (a) Low power response and (b) high power response. The frequency spectrum of each figure is seen in (c) and (d), respectively. In these figures the field magnitudes were normalized to the maxima of the incident electric fields. These figures also describe the bandwidth of the incident Gaussian pulse (see the red curves). }
\end{figure}

Such flexible absorbing performance can be seen with a broadband Gaussian pulse. This type of short pulse can be assumed to be closer to realistic high power threats to electronics from the viewpoint of the pulse width and frequency spectrum. The simulation results are shown in Fig.\ \ref{fig:gaussianPulse}, where a Gaussian pulse was modulated by a cosine function so that the resultant frequency spectrum is the red curve of Fig.\ \ref{fig:gaussianPulse} (c) (see Supplementary Material for more details on the pulse). In Fig.\ \ref{fig:gaussianPulse} (a), a low power Gaussian pulse illuminated the metasurface, resulting in 70 \% energy transmission with 3 \% reflection, i.e.\ only 27 \% absorption. The absorbing performance was markedly enhanced to 85 \%, when the input power was increased by a factor of 1,000 as in Fig.\ \ref{fig:gaussianPulse} (b). Comparison between Figs.\ \ref{fig:gaussianPulse} (c) and (d) reveals a significant reduction of the transmitted signal around 4.2 GHz, where our metasurface exhibited strong absorptance (see Fig.\ \ref{fig:highP} (c)). 

The absorbing performance of our metasurface can be further enhanced by increasing the thickness of the substrate. Note that the thickness was only about $\lambda$/50 here, representing a very thin metasurface. However, the metasurface still exhibited dependence on the incoming power level as well as on the waveform. In addition, in order to simplify the simulation and measurement process, we tested the metasurface in waveguides. This indicates that use of a signal source that generates only a surface wave can more clearly show the difference. Furthermore, we used schottky diodes in this study. However, the use of varactor diodes may enable us to sense the frequency of the incoming wave and tune the absorbing performance by varying the capacitance. We expect this would allow us to further improve the bandwidth of nonlinear metasurface.  

Use of rectification circuits can be seen in past studies \cite{auzanneau1998microwave, auzanneau1999artificial}, where diodes were used for antenna applications to rectify the incoming signal, resulting in, for example, extracting harmonics or behaving as an artificial magnetic conductor. However, for these purposes, it was not suitable to store the incoming energy temporarily and dissipate it, which is the key factor enabling the waveform dependence of the present structure.  

Our metasurface absorber effectively dissipated most of the energy of high power surface currents. This strong absorbing performance might appear similar to those of other metamaterial absorbers \cite{mtmAbsPRLpadilla, ultraThinAbs, My1stAbsPaper, cwFiltering,wakatsuchi2012performance}. However, while these metamaterial absorbers focused on absorption of waves illuminated from free space, our metasurface is specifically designed for high power surface currents. Nonetheless, our design is expected to respond to free space waves as well. The primary novelty of the nonlinear absorber described here is that it is the first surface designed specifically to respond to high power pulsed signals, thus representing the first waveform--dependent absorber. Additionally, the demonstrated metasurface was 162 mm long, which is about two wavelengths at 4 GHz near the absorption peak. This indicates that applying the nonlinear metasurface around only a two--wavelength radius from sensitive electronics may be enough to protect them from high power threats while still allowing them to communicate with external signal sources. 

We have introduced a new kind of nonlinear metasurface that is designed for absorption for high power microwave pulses. This surface works by rectifying the incoming signal using diodes, and storing the static field in capacitors, to be dissipated in resistors between pulses. Thus, the surface responds to not only the power level, but also the temporal profile of the incoming signal in order to specifically absorb short, high power microwave pulses. 

This work was supported by the Office Naval Research under Grant N00014--11--1--0460. 



\providecommand{\noopsort}[1]{}\providecommand{\singleletter}[1]{#1}%

\clearpage
\appendix
{\Large \textbf{Supplemental Material}}

\begin{table*}[b!]
\caption{Fourier component of various modes for half and full wave rectifications}
\label{tab:rect}
\begin{center}
\begin{tabular}{c|p{3cm}p{3.8cm}p{3.8cm}}
Fourier term & W/O rectification: & Half wave rectification: & Full wave rectification:\\
 & $\cos$ & ($\cos$+$|\cos|$)/2 & $|\cos|$\\
\hline
0&0&1/$\pi$&2$\pi$\\
$f$ &1/2&1/4&0\\
2$f$ &0&1/(3$\pi$)&2/(3$\pi$)\\
3$f$ &0&0&0\\
4$f$ &0&-1/(15$\pi$)&-2/(15$\pi$)\\
\hline
\end{tabular}
\end{center}
\end{table*}

\textbf{\begin{itemize}
\item Rectification
\item Simulation method
\item Measurement method
\item Rectification to a static field 
\item High power absorbing performance 
\item Gaussian pulse performance 
\end{itemize}
}
\section*{Rectification}
Microwave diodes rectify incoming signals to a static field in the following manner. Surface waves can be represented by a cosine function $\cos{(2\pi ft)}$, where $f$ and $t$ are respectively the frequency and time. Note that for simplicity other variables including the spatial position, phase delay, etc, are all omitted here. When the input signal is fourier--transformed (i.e.\ $\int^{\infty} _{-\infty}\cos{(2\pi ft)}e^{-j2\pi t}dt$), the output spectrum contains only $f$. If the surface current is rectified by a diode, however, the rectified signal becomes ($\cos{(2\pi ft)}+|\cos{(2\pi ft)}|$)/2, which through a fourier transform results in an infinite set of frequencies with decreasing magnitude. The largest term is at zero frequency, or a static field. This rectification to a static field is further enhanced, if a full wave rectification is introduced as the metasurface demonstrated in the paper. In this case the incoming signal is rectified to $|\cos{(2\pi ft)}|$. All of these are summarized in Table \ref{tab:rect}. 

\section*{Simulation method}
All the simulations were performed by electromagnetic/circuit co--simulation. First, as drawn in Fig.\ \ref{fig:models} (e), we calculated the scattering parameters through an electromagnetic simulator Ansys HFSS 15.0. The metasurface simulated here had lumped ports which were later connected to circuit components, such as diodes, in circuit simulations performed by a circuit simulator Ansoft Designer 8.0. Effectively, this is equivalent to directly connecting them to the metasurface in the electromagnetic simulation. Hence, all the scattering parameters and absorptance were calculated by the circuit simulator. 

\begin{figure}[t!]
\includegraphics[width=0.45\textwidth]{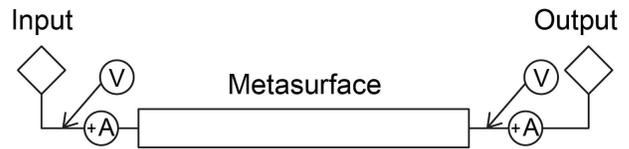}
\caption{\label{fig:circuitSim} Circuit simulation. For the actual model we also connected lumped ports to the metasurface. }
\end{figure}

\begin{figure}[t!]
\includegraphics[width=0.25\textwidth]{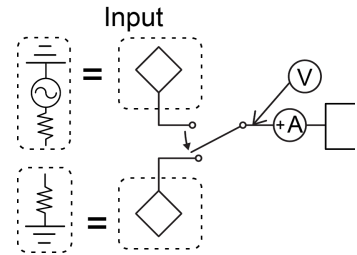}
\caption{\label{fig:switch} Introduction of switching system to produce pulses. In an arbitrary time the metasurface model was disconnected from the original input port and connected to a new port which did not have any excitation source. Each port had the same impedance as the port impedance of the metasurface model. }
\end{figure}

The performance of high power pulse and CW simulations were both evaluated in time domain using the electric circuit shown in Fig.\ \ref{fig:circuitSim}, where for simplicity we abbreviated lumped ports. In these simulations the input power was theoretically estimated by 
\begin{eqnarray}
P_{in}(t)=2P_{0}\sin^2{(2\pi ft)},\label{eq:PinCircuit}
\end{eqnarray}
\\where $P_{in}$ and $P_{0}$ are, respectively, the instant power at the input port and the magnitude of the input power. Since the voltage and current meters next to the input port read a power $P_{m_1}$ combining both incident and reflected powers, the reflected power was calculated by subtracting $P_{m_1}$ from eq.\ (\ref{eq:PinCircuit}). For the transmitted power, the meters next to the output port were used, since there is no signal from the output port.  

For the pulse simulations, a switching system was inserted between the input port and neighboring meters as Fig.\ \ref{fig:switch}. This system played a role in switching the connection between the metasurface and input port. First, the metasurface model was connected to the original input port which generated a CW signal. Then, the port was disconnected and the
metasurface was connected to a new port, which did not have any signal source. As a result, an arbitrary width of pulse was produced. To evaluate the absorptance, the reflected and transmitted energies were integrated over time and divided by the input energy. These values were subtracted from one later. 

The high power CW simulations were more straightforward, i.e.\ the simulation was performed until a steady state, where the reflected and transmitted power were averaged and divided by the input power. There were a couple of factors to lead to instability in the simulations, for example, due to passivity and use of commercial diode models with small signals. For this reason some simulations were truncated earlier than others to avoid diverged results. 

\section*{Measurement method}
High power measurements were performed using the measurement system sketched in Fig.\ \ref{fig:models} (f). An input signal was generated from a signal generator (Agilent N5181A) and passed through an amplifier (OPHIR 5193). The amplified signal was coupled into waveguides through an isolator to protect the amplifier from possible reflections. In addition, directional couplers (KRYTAR 102008010) were used to monitor the incident and reflected signals, which were measured by power sensors (Agilent N1921A). The power sensors were protected from high power signals using an attenuator. In order to measure the scattering parameters as accurately as possible, the high power measurements were performed after several calibration steps. Some of the measurement devices were automatically controlled through LabVIEW software such that the input power and frequency could be quickly swept, which facilitated the measurement process substantially. A pulse width and duty cycle were changed with the signal generator and LabVIEW. 

Extra cares must have been paid on the pulse measurements. Power meters used (Agilent N1911A) had some operating modes, e.g.\ averaging mode and peak mode. Since our simulation results showed the same peaks for different pulse widths in time domain, we adopted the averaging mode. However, this indicates that depending on the duty cycle of the pulse, we needed to offset the received signals, e.g.\ if the received signal is 10 dBm and the duty cycle is 10 \% (i.e.\ -10 dB), then the actual magnitude is 20 dBm (i.e.\ 10 dBm - (-10 dB)). Due to the noise floor of the power meters, the duty cycle could not be set too small, otherwise the measured value would read the noise floor. This issue, however, needed to be compromised with another problem, i.e.\ the discharging time of the stored energy. Since the signal generator repeatedly produced pulses, the duty cycle was set long enough to ensure fully discharging the electric charges stored in capacitors. Because of these two issues we decided to set the duty cycle to 1 \% after some test measurements. 

Another point to note here is that the reflection was significantly small in the pulse measurements. Hence, the absorptance was estimated from the transmittance only, i.e.\ from 1 - $T$, where $T$ represents the transmittance. 

\begin{figure}[t!]
\includegraphics[width=0.45\textwidth]{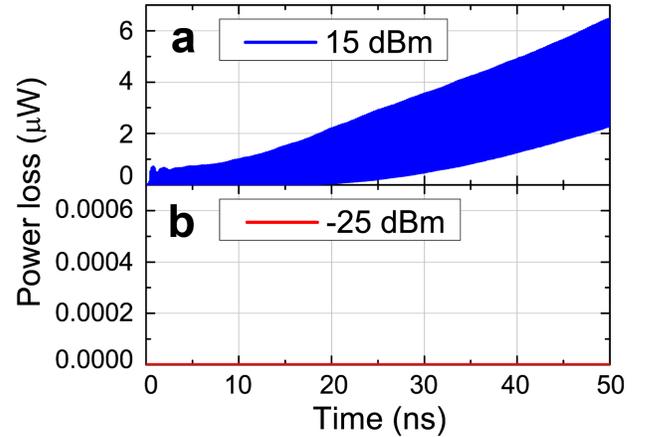}
\caption{\label{fig:rect} Numerical proof of rectification to a static field. Power loss at the second resistor from the front periodic unit was calculated with (a) 15 dBm and (b) -25 dBm pulses at 4.2 GHz. }
\end{figure}

\begin{figure*}[t!]
\includegraphics[width=0.95\textwidth]{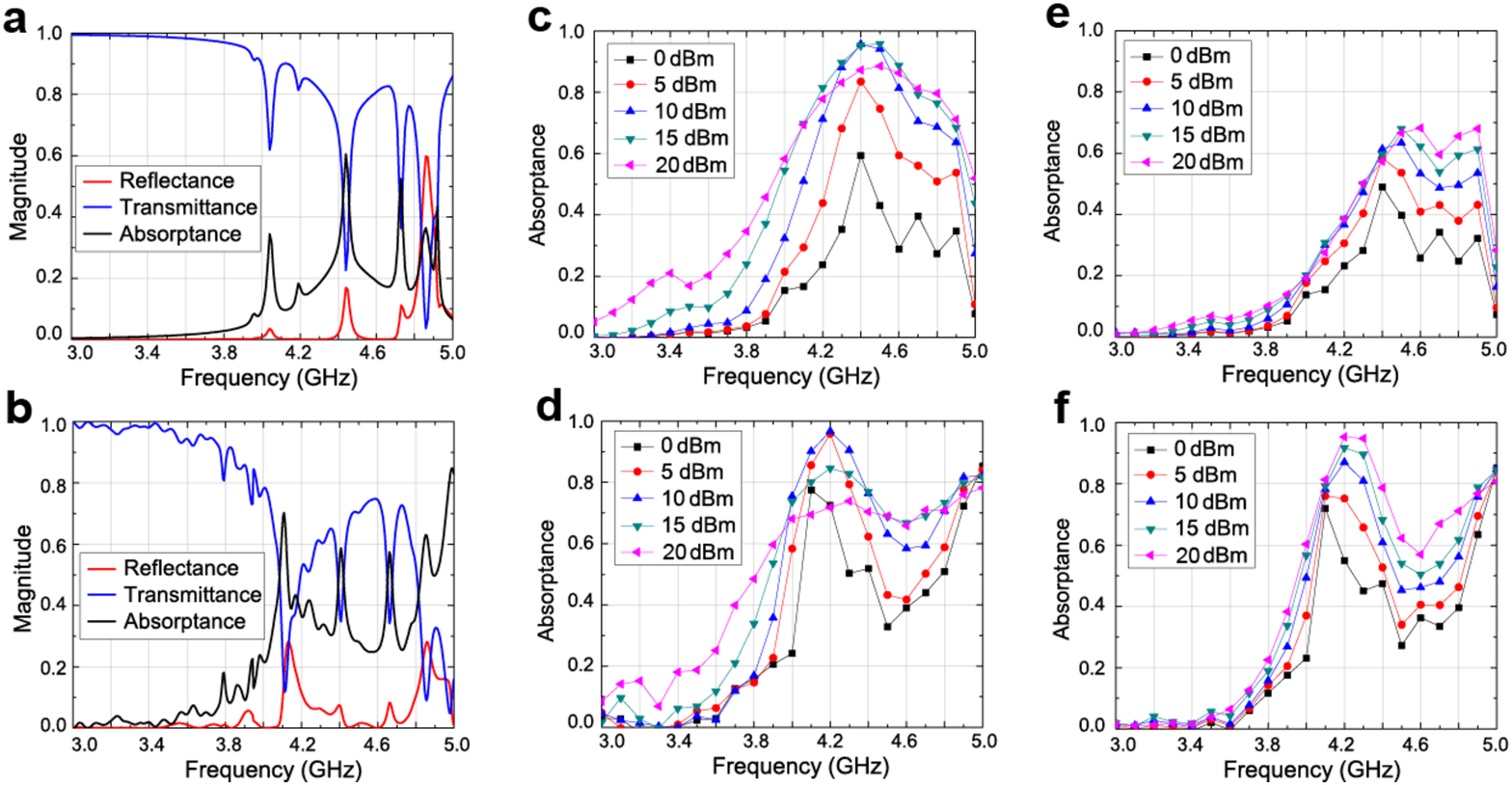}
\caption{\label{fig:1kohm} Scattering and absorbing profile with 1 k$\Omega$ resistors. Simulation results are plotted in the left figures (i.e.\ (a), (c) and (e)), while measurement results in the right ((b), (d) and (f)). Low power CW responses are shown in (a) and (b). High power pulse absorptances are shown in (c) and (d), while high power CW absorptances in (e) and (f). In (c) and (d) the pulse width was set to 50 ns. }
\end{figure*}

\begin{figure*}[t!]
\includegraphics[width=0.95\textwidth]{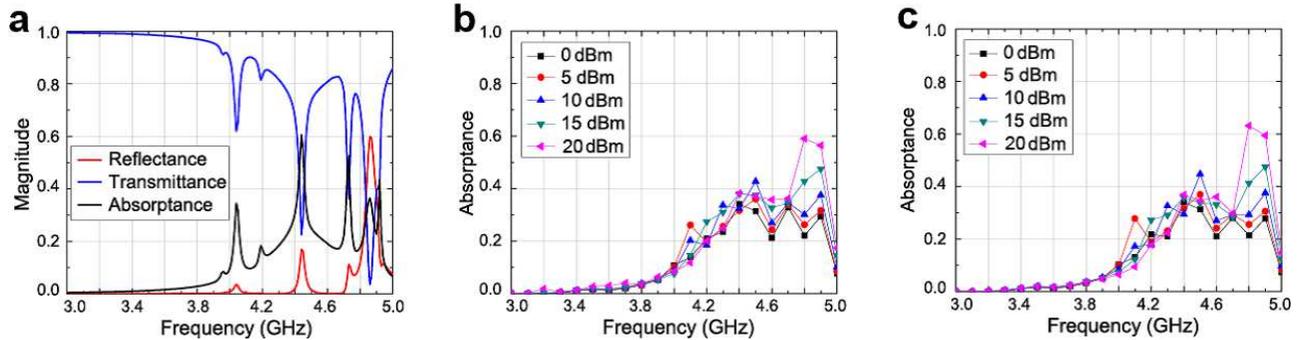}
\caption{\label{fig:woCaps} Simulation results of scattering and absorbing profile without capacitors (see Fig.\ \ref{fig:pulseWidthDep} (c) too). (a) Low power CW response. (b) High power pulse absorptances. (c) High power CW absorptances. In (b) the pulse width was set to 50 ns. }
\end{figure*}

\section*{Rectification to a static field}
We numerically verified that diodes rectify incoming signals to a static field. As an example, Fig.\ \ref{fig:rect} plots curves of the instant power loss at the second resistor from the front unit with two different power levels, i.e.\ 15 and -25 dBm at 4.2 GHz. If a normal cosine function of current flows into a resistor, then the instant power loss follows a square of the cosine function. As a result, the curve regularly touches the zero watt line. With the rectification, however, the curve moves away from the zero watt as in Fig.\ \ref{fig:rect} (a), since the nonlinearity added a static field to the curve. In our structure, if the voltage difference across a diode is not large enough to turn on the diode, then there is no current flowing into the resistor (cf.\ Fig.\ \ref{fig:models} (a)). Hence, the resultant power loss becomes as Fig.\ \ref{fig:rect} (b). 

\section*{High power absorbing performance}
The absorbing performance to high power CW signals can be controlled by varying the resistance value. As mentioned earlier, Fig.\ \ref{fig:highP} used 10 k$\Omega$ for the resistors. When the resistance was reduced to 1 k$\Omega$, the high power CW performance appeared as Fig.\ \ref{fig:1kohm} and approached the pulse performance. This is because the reduction of the resistance allowed direct dissipation of the rectified energy from the diodes to the resistor. 

In Fig.\ \ref{fig:pulseWidthDep} (c) a pulse width dependence of a metasurface without capacitors was shown. This structure exhibits almost no dependence on both power and waveform as seen in Fig.\ \ref{fig:woCaps}, since there is no capacitor to temporarily store the pulse energy, even if the incoming signal is rectified by diodes. Also, the resistance used (i.e.\ 10 k$\Omega$) is also large enough to prevent direct energy dissipation from the diodes. 

\section*{Gaussian pulse performance}
We applied the following equation to the Gaussian pulse used in the paper: 
\begin{eqnarray}
V_{in}(t)=V_0(\exp{(-\alpha (t -t _0)^2)})\cos{(\omega _c(t -t _0))},
\end{eqnarray}
where $V_{in}$ and $V_0$ respectively denote the instant input port voltage and a scaling factor. The other parameters were $\alpha=(2/t_0)^2$, $\omega _c=10\pi/t_0$ and $t_0=1/(0.28 \cdot 10^9\pi)$. This function was discretized fine enough in time domain and used as a piecewise function of an input source in Ansys Designer. Under these circumstances, the Gaussian pulse had strong frequency components between 4 and 5 GHz (i.e.\ the red curve of Fig.\ \ref{fig:gaussianPulse} (c)), where our metasurface exhibited strong absorbing performance (Fig.\ \ref{fig:highP} (c)). 

\begin{figure}[b!]
\includegraphics[width=0.45\textwidth]{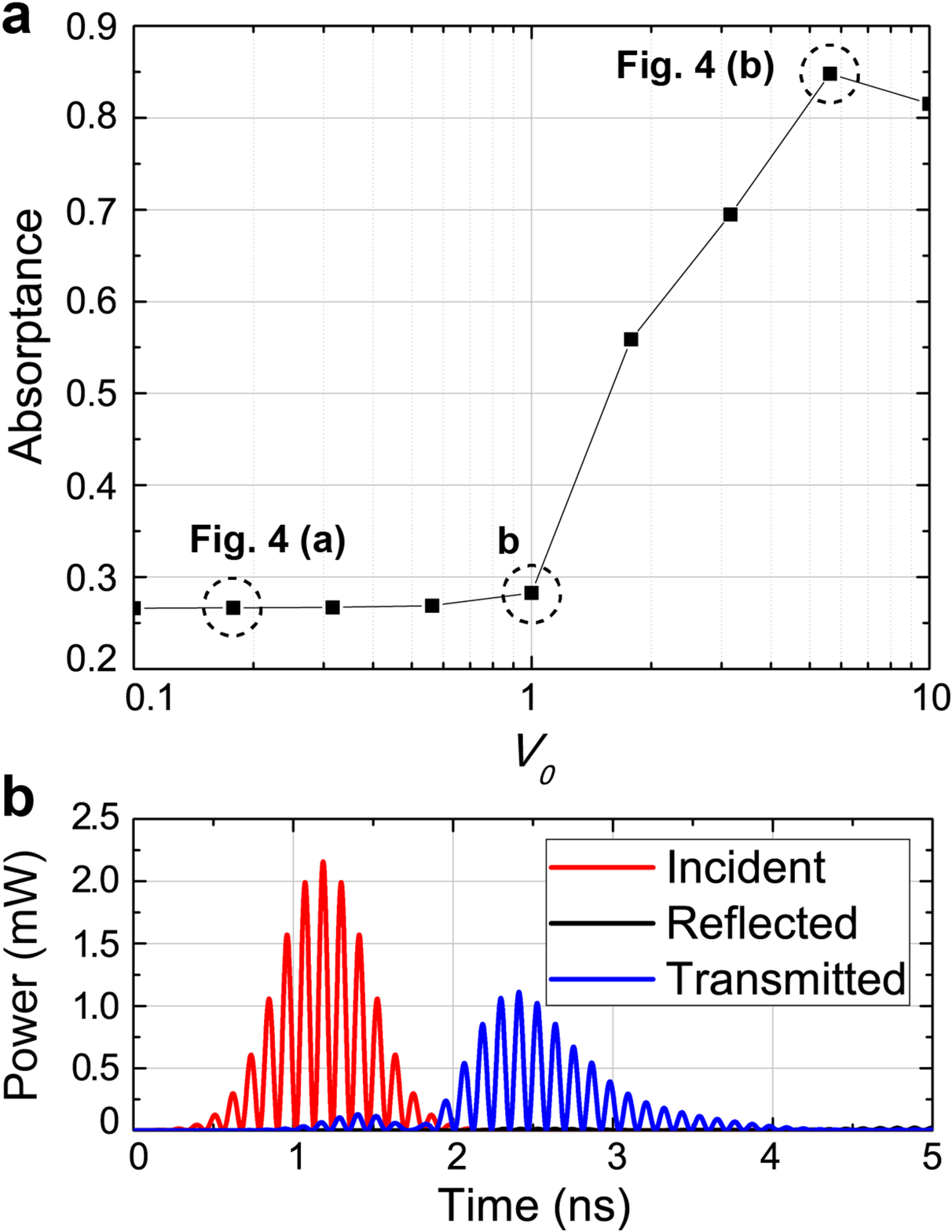}
\caption{\label{fig:GaussDependence} More details on Gaussian pulse response. (a) The dependence on scaling factor $V_0$. (b) The incident, reflected and transmitted powers at $V_0=1$.}
\end{figure}

More details on the voltage dependence are described in Fig.\ \ref{fig:GaussDependence} (a), where the absorbing performance is shown as a function of $V_0$. In this figure the absorbing performance started from 27 \% and, as $V_0$ increased, reached 85 \%. Especially, the absorptance significantly varied between $V_0=1$ and $V_0 \sim 5.6$. Here each of the maximum instant powers corresponded to approximately 2 and 60 mW, respectively, as seen in Figs.\ \ref{fig:GaussDependence} (b) and \ref{fig:gaussianPulse} (b). It turned out from these figures and Fig.\ \ref{fig:highP} (c) that this is the same range as where the metasurface varied the absorbing performance toward single frequency pulses, although the single frequency pulses had a longer pulse width (i.e.\ 50 ns). 

\end{document}